\documentclass{optica-article}

\journal{opticajournal} % for journals or Optica Open

\articletype{Research Article}

 \usepackage{lineno}
\usepackage{siunitx}
\DeclareSIUnit\dBm{dBm}
\usepackage{comment}
\begin{document}

\title{Semi-classical analysis of noise squeezing in Kerr microresonators}

\author{Sara Persia,\authormark{1,*,$\dagger$} Lev Fandeev,\authormark{2,$\dagger$} Fuchuan Lei,\authormark{3,$\dagger$} 
Bastian Ruhnke,\authormark{2} Alexander E. Ulanov,\authormark{2} Vaishali Adya,\authormark{4} Victor Torres-Company,\authormark{1,**} and Tobias Herr\authormark{2,5,***}}

\address{\authormark{1}Department of Microtechnology and Nanoscience (MC2), Photonics Laboratory, Chalmers University of Technology, SE-41296 G\"oteborg, Sweden\\
\authormark{2}Deutsches Elektronen-Synchrotron DESY, Hamburg, Germany\\
\authormark{3}State Key Laboratory of Integrated Optoelectronics, School of Physics, Northeast Normal University, 130024 Changchun, China\\
\authormark{4}Department of Applied Physics, KTH Royal Institute of Technology, SE-106 91 Stockholm, Sweden\\
\authormark{5}Physics Department, University of Hamburg UHH, Hamburg, Germany\\
\authormark{$\dagger$}These authors contributed equally to this work.\\}  

\email{\authormark{*}persias@chalmers.se \authormark{**}torresv@chalmers.se \authormark{***}tobias.herr@desy.de} %% email address is required

\begin{comment}
\author[1,$\dagger$]{Sara Persia}
\author[2,$\dagger$]{Lev Fandeev}
\author[3,$\dagger$]{Fuchuan Lei}
\author[2]{Bastian Ruhnke}
\author[2]{Alexander E. Ulanov}
\author[4]{Vaishali Adya}
\author[1*]{Victor Torres-Company}
\author[2,5**]{Tobias Herr}

\affil[1]{Department of Microtechnology and Nanoscience (MC2), Photonics Laboratory, Chalmers University of Technology, SE-41296 G\"oteborg, Sweden}
\affil[2]{Deutsches Elektronen-Synchrotron DESY, Hamburg, Germany}
\affil[3]{State Key Laboratory of Integrated Optoelectronics, School of Physics, Northeast Normal University, 130024 Changchun, China}
\affil[4]{Department of Applied Physics, KTH Royal Institute of Technology, SE-106 91 Stockholm, Sweden}
\affil[5]{Physics Department, University of Hamburg UHH, Hamburg, Germany}

\affil[$\dagger$]{These authors contributed equally to this work.}

%\affil[*]{[persias@chalmers.se]}
\affil[*]{[torresv@chalmers.se]}
\affil[**]{[tobias.herr@desy.de]} 

%\email{\authormark{*}persias@chalmers.se \authormark{**}torresv@chalmers.se \authormark{***}tobias.herr@desy.de} %% email address is required
\end{comment}

\begin{abstract*}
Kerr microresonators are integrated nonlinear devices widely used to generate classical optical frequency combs and nonclassical states of light. In particular, they provide a promising platform for squeezed light sources, which are important elements of emerging quantum sensing, communication, and computing technologies. Designing such sources requires numerical tools that efficiently incorporate realistic device physics and noise processes. Here, we show that established semi-classical frameworks based on the Lugiato-Lefever equation and the Ikeda map, instrumental in the development of classical light sources, can also predict the quadrature-noise properties of squeezed light when vacuum fluctuations are properly included. We consider three canonical configurations producing two-mode quadrature squeezing, single-mode quadrature squeezing, and intensity difference squeezing. In each case, the semi-classical simulations produce quadrature variances and noise spectra in close agreement with the quantum theoretical predictions. These frameworks therefore provide practical tools for designing squeezed light sources, with the capability to incorporate additional noise mechanisms, nonlinear effects, and complex resonator arrangements. More fundamentally, our results show that key experimental signatures of squeezing can be modeled as phase-sensitive amplification of properly normalized vacuum fluctuations within stochastic semi-classical field equations. This observation helps distinguish squeezing properties accessible through semi-classical field evolution from those intrinsically requiring a full quantum description.
\end{abstract*}

% \begin{document}

% \maketitle

% use {asbstract*} to suppress the copyright line. Copyright information will be added in production

%%%%%%%%%%%%%%%%%%%%%%%%%%  body  %%%%%%%%%%%%%%%%%%%%%%%%%%

\section{Introduction}
\begin{figure*}[!]
\centering\includegraphics[]{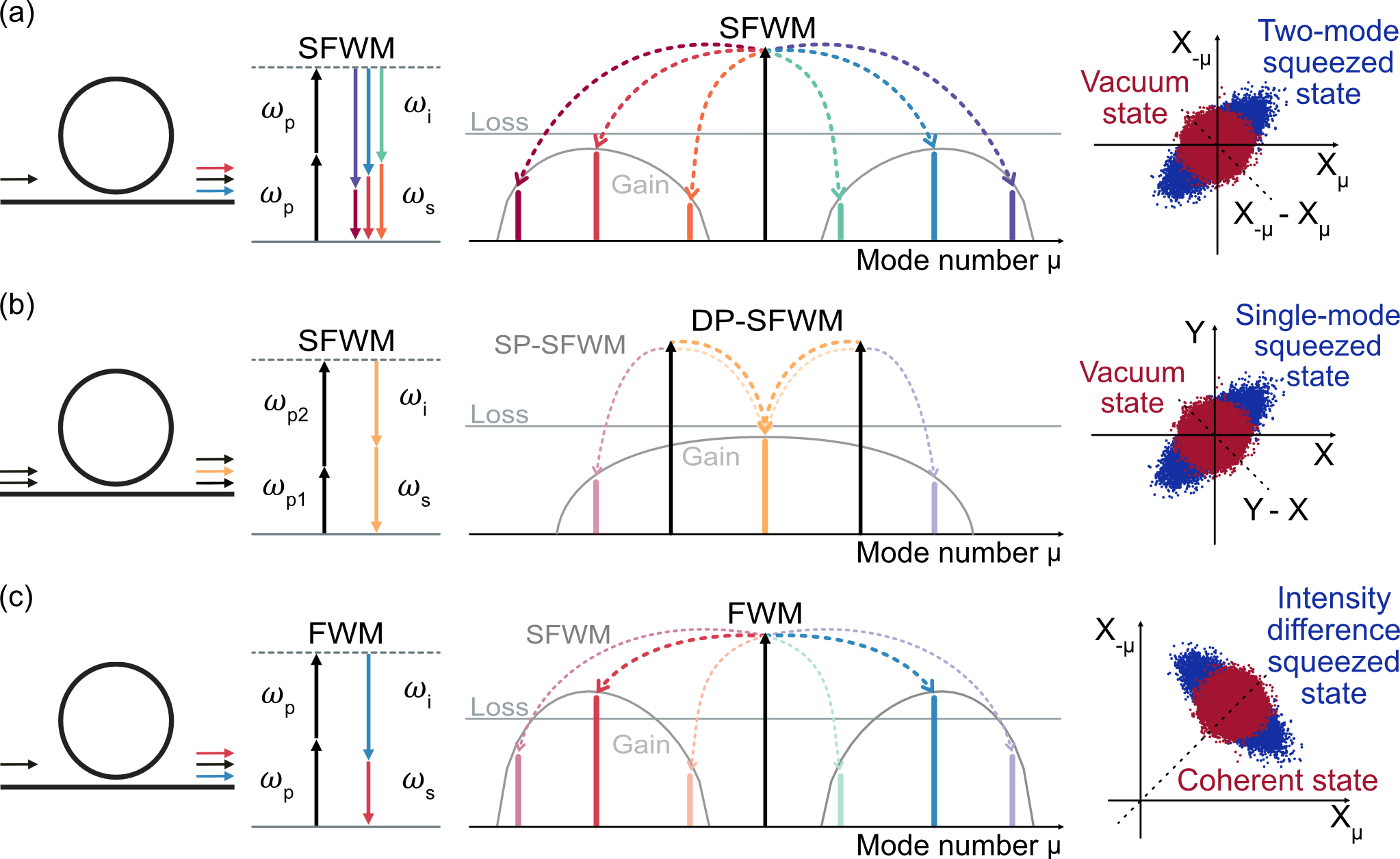}
\caption{Schematic of the generation of (a) a two-mode quadrature squeezed state, (b) a single-mode quadrature squeezed state, and (c) an intensity difference squeezed state in a Kerr microresonator. For each case, we show the relation between the pump(s), signal, and idler frequencies characterizing the FWM processes and the corresponding output frequency spectra. Here, pumps are shown as black arrows, colorful continuous lines represent the generated modes, dashed arrows denote the nonlinear interactions, and light, thinner arrows represent unwanted nonlinear processes. The phase space representations of the squeezed states (blue) compared with the vacuum or coherent states (red) are also shown.}
\label{fig1}
\end{figure*}
Kerr microresonators are key components in photonic integrated circuits (PICs), especially in applications exploiting optical nonlinearities \cite{Chang2022}. 
%They typically consist of a straight bus waveguide placed close to a looped waveguide that acts as an optical cavity \cite{ring}. 
In the classical regime, these devices are often used to generate optical frequency combs, consisting of multiple, evenly spaced, coherent frequency modes \cite{DelHaye2007, Pasquazi2018, herr2026}. Their spectral properties are of great interest for applications including optical telecommunications \cite{Marin-Palomo2017}, spectroscopy \cite{dualcomb}, and optical atomic clocks \cite{Wu2025}. These microcombs are typically generated through four-wave mixing (FWM) of frequency components when the input pump power exceeds the characteristic optical parametric threshold $P_\text{th}$, at which the parametric gain exceeds the total cavity loss, allowing multiple frequency modes to be amplified \cite{Herr}. 
The classical dynamics of Kerr microresonators has been extensively investigated both experimentally and theoretically. Well-established semi-classical models, like those based on the Lugiato-Lefever equation (LLE) \cite{LLE, haelterman1992DissipativeModulationInstability} and its coupled-mode equation (CME) formulation \cite{PhysRevA.82.033801}, as well as the Ikeda map \cite{Ikeda1979}, have been developed and refined to explore the practical limits of these devices \cite{PhysRevA.87.053852, Coen:13, Hansson:15, Lei2022, Helgason2021, Murdoch:26, herr2026}.  

Recent works have also shown the potential of integrated Kerr microresonators for quantum photonics \cite{PICQ, Kues2019, Quesada2022}. Quantum states of light, such as squeezed states, have been widely generated in these devices \cite{Dutt2015, Dutt:16, Zhao2020, Zhang2021, Vaidya, Yang2021, Jahanbozorgi2023, Ulanov2025, shen2025StrongNanophotonicQuantum, Persia:26}. These sources could support applications in quantum sensing \cite{Lawrie2019, GW1, GW2, wang20252}, quantum communication \cite{pirandola2020AdvancesQuantumCryptography}, and continuous-variable quantum computing \cite{ madsen2022QuantumComputationalAdvantage, AghaeeRad2025}. Such states are typically generated through spontaneous FWM (SFWM) by pumping the device below $P_\text{th}$, where, classically, the pump is the only oscillating mode in the cavity. From a quantum perspective, two pump photons are instead converted into photons at different frequencies, effectively amplifying the vacuum fluctuations populating the other modes \cite{Kues2019,Yang2021, Guidry2022,jia2025ContinuousvariableMultipartiteEntanglement} . Due to the growing interest in employing these versatile devices in complex quantum applications, a deep understanding of their quantum dynamics is crucial. So far, the quantum dynamics of microresonators has been described almost exclusively within the quantum-mechanical framework \cite{lugiato, Chembo2016, Guidry2022, Quesada2022, Guidry2023, Lustig2025, Sloan2025, steady, Kim, Dean2026}.  % Significant effort has gone into developing a complete quantum formalism capable of describing and predicting the quantum behavior of Kerr microresonators, including significant progress in describing the origin of quantum frequency combs \cite{Guidry2022}, or the emergence of squeezed states in complex Kerr soliton states \cite{Guidry2023, Lustig2025}.

Here we show that the classical models of LLE, CME and Ikeda map can be extended to predict numerically the noise properties of squeezed light generated in Kerr microresonators. Following earlier work describing quantum noise in mode-locked lasers \cite{haus2000, Paschotta2004} and microresonators \cite{bao2021}, we supplement the classical field equations with properly normalized vacuum noise. We describe this approach for both the Ikeda map and the coupled-mode LLE, and validate it against the analytical quantum predictions for (i) two-mode quadrature squeezing \cite{steady}, (ii) single-mode quadrature squeezing \cite{Chembo2016, Guidry2023}, and (iii) intensity-difference squeezing \cite{Chembo2016}. The simulations give direct access to the evolution of all cavity modes, from which quadrature distributions and noise spectra can be extracted. Because the classical models already routinely describe additional effects such as higher-order dispersion, cascaded four-wave mixing, Raman and thermorefractive effects, or multi-resonator geometries, the extension presented here allows them to serve as practical design tools for integrated squeezed-light sources in the same way they serve for classical microcombs.

% These examples show that, so far, the quantum dynamics of microresonators has been primarily described in the quantum mechanical framework, while semi-classical models have been employed mainly to analyze the classical regime. In this work, however, we show that the generation of squeezed states in integrated Kerr microresonators can be accurately described and predicted with the same semi-classical models typically used to investigate the creation of microcombs. In particular, we demonstrate the use of the LLE and the Ikeda map to simulate three canonical quantum properties of light: two-mode quadrature squeezing \cite{steady}, single-mode quadrature squeezing \cite{Ulanov2025}, and intensity difference squeezing \cite{Chembo2016}. The numerical simulations provide fast and direct access to the temporal evolution of the frequency modes, whose properties are analyzed in the phase space and through the corresponding noise spectra. The results show good agreement with the corresponding quantum descriptions, thereby validating our approach. These outcomes surprisingly reveal that semi-classical formalisms can be used as additional tools for investigating squeezing generation in integrated nonlinear photonics.

\section{Squeezed states in Kerr microresonators}
Squeezed states are quantum states of light exhibiting noise reduction of one specific quadrature of an optical field below the vacuum noise level (or shot noise), at the expense of increased noise in the conjugate quadrature, as required by the Heisenberg uncertainty principle \cite{Lvovsky}. Their generation is based on nonlinear interaction processes that create correlations between the quantum fluctuations of vacuum fields \cite{Chembo2016}. In Kerr microresonators, this interaction is provided by the FWM arising from the third-order nonlinear susceptibility $\chi^{(3)}$ of the waveguide material, which, without loss of generality, is assumed to be SiN throughout this work. %Moreover, to achieve high levels of squeezing, the device needs to be strongly overcoupled, which means having low intracavity losses and a high extraction rate \cite{Chembo2016, Quesada2022}. 
Fig. \ref{fig1} summarizes the three types of squeezed states considered in this study.

Two-mode quadrature squeezing is typically generated between two cavity modes, usually called signal and idler, from a single pump of power $ P_\text{in}$ below the oscillation threshold $\left ( P_\text{in} < P_\text{th} \right)$. As shown in Fig. \ref{fig1}(a), SFWM converts two pump photons into two photons of lower and higher energy, satisfying the energy conservation relation $2\hbar \omega_p = \hbar \omega_s + \hbar \omega_i$, where $\hbar$ is the Planck constant divided by $2\pi$, and $\omega_{p(s,i)}$is the angular frequency of the pump (signal and idler) mode. The resulting modes share quantum correlations, forming a two-mode quadrature squeezed state whose noise reduction is therefore a collective property of the mode pair. In phase space, the characteristic squeezed noise ellipse arises only when the signal and idler quadratures are combined and analyzed together \cite{Caves85}. In this pump configuration, energy and momentum conservation of the nonlinear process enable the simultaneous generation of several two-mode squeezed vacuum states \cite{Kues2019, Yang2021, Guidry2022}, with the strongest noise reduction obtained for the modes corresponding to the peaks of the parametric gain spectrum \cite{steady}.

A second type of squeezing commonly generated in a Kerr microresonator is single-mode quadrature squeezing. As shown in Fig. \ref{fig1}(b), it can be generated by pumping the resonator with two input fields at different frequencies $\omega_{p1} \ne \omega_{p2}$. In this dual-pump spontaneous FWM (DP-SFWM) process, one photon from each pump mode is converted into two degenerate photons in the central signal mode, according to the energy relation $\hbar \omega_{p1}  + \hbar \omega_{p2}  =  2 \hbar \omega_s $. However, this pump configuration is affected by additional nonlinear interactions, such as single-pump spontaneous FWM (SP-SFWM) and Bragg-scattering FWM, which introduce extra noise and degrade the achievable squeezing \cite{Seifoory2022}. Several techniques have been proposed to suppress these unwanted processes \cite{Zhang2021, Viola2024, Ulanov2025}. Consequently, most quantum models consider a simplified three-mode system consisting only of the two pumps and the central signal \cite{Zhao2020, Zhang2021}. 

The third type of quantum state considered in this work is the intensity difference squeezed state, illustrated in Fig. \ref{fig1}(c). It is obtained by pumping the microresonator with a single pump above threshold $\left(P_\text{in} > P_\text{th}\right)$. In this regime, the FWM process generates bright signal and idler modes with power sufficiently larger than vacuum fields. As for two-mode squeezed states, noise squeezing is a shared property of the two bright modes. Here, the photon number (intensity) difference between the signal and the idler modes is the squeezed quantity. It arises from the same process that produces two mode squeezing below threshold \cite{Chembo2016}. 

% The squeezed states described above form the main building blocks for several quantum applications, such as Gaussian boson sampling \cite{Arrazola2021}, quantum photonic computing \cite{madsen2022QuantumComputationalAdvantage,AghaeeRad2025}, quantum key distribution \cite{pirandola2020AdvancesQuantumCryptography}, quantum sensing \cite{Lawrie2019, wang20252}, and quantum microscopy \cite{Casacio2021}. Their generation has been widely demonstrated in microresonators \cite{Dutt2015,Dutt:16, Zhao2020, Zhang2021, Vaidya, Yang2021, Jahanbozorgi2023, Ulanov2025, shen2025StrongNanophotonicQuantum,Persia:26}, highlighting the potential for scalability, while their properties have been theoretically analyzed using various quantum frameworks\cite{lugiato, Chembo2016, Guidry2022, Quesada2022, Guidry2023, Lustig2025, Sloan2025, steady, Kim, Dean2026}.

\section{Semi-classical simulation frameworks}
In this section, we describe how vacuum noise is incorporated into the semi-classical models of Kerr microresonators, whose full formulations can be found in Refs. \cite{Agrawal2012,PhysRevA.87.053852, Hansson}. 
For clarity, we restate the set of equations used to obtain the results shown in the next section, and present our analysis of the noise terms.

Several theoretical models have been developed to describe the nonlinear dynamics of Kerr microresonators in the classical regime and are regularly used to simulate frequency comb generation \cite{Hansson, herr2026, murdoch2026}. Besides simplifying assumptions, a key difference between the available models is the perspective from which the system is analyzed, either in the time domain through a single partial differential equation (PDE) \cite{PhysRevA.87.053852}, or in the frequency domain through coupled differential equations that describe the temporal evolution of each spectral component of the field (modal expansion approach) \cite{PhysRevA.82.033801}. Both formulations can be implemented numerically and provide equivalent results \cite{PhysRevA.87.053852}. In this work, we use both approaches to simulate noise properties of squeezed light generated in the processes shown in Fig. \ref{fig1}.

The first model we implement is based on the Ikeda map, a time-domain approach that combines a PDE, describing the variation of the intracavity field envelope $A_n$ during a single roundtrip $n$ in the cavity, with the input-output relations of the ring-bus waveguide coupling region. The field evolution is described by the modified nonlinear Schrödinger equation in the reference frame moving at the group velocity defined as \cite{Agrawal2012}
\begin{equation}
    \frac{\partial A_n(z,t)}{\partial z} = \left[ -\frac{\alpha}{2} +i\beta_0 - \frac{i\beta_2}{2}\frac{\partial^2 }{\partial t^2} + i\gamma|A_n|^2 \right] A_n(z,t)
    \label{nlse}
\end{equation}
where $z$ is the propagation direction, $t$ is the fast time variable, $\beta_2$ is the group velocity dispersion, $\gamma$ is the nonlinear coefficient, and $\beta_0= -\delta_0 T_R/L$ accounts for the accumulated linear phase shift depending on the cavity length $L$ and the detuning $\delta_0 = \omega_0 - \omega_p$ between the pump laser frequency and the cold cavity resonance $\omega_0$, with $T_R$ the roundtrip time.
The propagation loss $\alpha$ is related to the intrinsic quality factor $Q_i = \omega_p n_g L\sqrt{a}/[2(1-a)c]$ through the optical power attenuation per roundtrip $a^2 = e^{-\alpha L}$, with $n_g$ being the group index, and $c$ the speed of light in vacuum. Here, we normalize the field such that $|A|^2 $ is the field optical power $P$. The cubic term in Eq. (\ref{nlse}) describes the nonlinear Kerr interactions responsible for FWM. We implement this equation numerically using the split-step method, alternating linear and nonlinear propagation over small steps. The coupling equations relating the cavity output $A_\text{out}$, input $A_\text{in}$, and intracavity fields of subsequent roundtrips are
\begin{equation}
    A_\text{out} = A_\text{in}\sqrt{1-\theta} - A_n(L,t)\sqrt{\theta}
    \label{Aout}
\end{equation}
\begin{equation}
    A_{n+1}(0,t) = A_\text{in}\sqrt{\theta} + A_n(L,t)\sqrt{1-\theta}.
    \label{An1}
\end{equation}
Here, $\theta$ is the ring-bus coupling coefficient related to the extrinsic quality factor as $Q_e = \omega_p n_g L\sqrt{r}/[2(1-r)c]$, with $|r| = \sqrt{1-\theta}$.

The Ikeda map defined by Eqs. (\ref{nlse})-(\ref{An1}) can be reduced to the LLE under the mean-field approximation, where the intracavity variation from one roundtrip % of duration $T_\mathrm{R}$ 
to the next is small and the resonator-to-bus coupling is weak. In this description, the field is expressed in a co-moving frame as $A'(\phi,t)$ where the angular coordinate $\phi\in [0,2\pi)$ spans one resonator roundtrip; we normalize $A'$ such that $|A'|^2 = P T_\mathrm{R}/(\hbar \omega)$ is the intracavity photon number. The resulting equation describing the propagation over many roundtrips is \cite{PhysRevA.87.053852}
\begin{equation}
    \frac{\partial A'( \phi,t)}{\partial t} = -\left ( \frac{\kappa}{2} +i\delta_0 \right)A' + \frac{i D_2}{2}\frac{\partial^2 A'}{\partial \phi^2 } + ig_0|A'|^2 A' + \sqrt{\kappa_\mathrm{ex}} s_\text{in}
    \label{LLE}
\end{equation}
where $\kappa = \kappa_0 + \kappa_\mathrm{ex}$ is the total resonator decay rate combining the internal decay rate
$\kappa_0 = \alpha L/T_\mathrm{R}$ and resonator-bus coupling rate $\kappa_\mathrm{ex}= \theta/T_\mathrm{R}$, $g_0 = \hbar \omega_p v_g^2 \gamma/L$ is the nonlinear coupling coefficient, with $v_g$ the group velocity, $D_2 = -\text{FSR}^3L\beta_2(2\pi)^2$ is the second-order dispersion, FSR is the free spectral range, and $|s_\mathrm{in}|^2 = P_\mathrm{in}/(\hbar \omega_p)$ denotes the normalized pump field.

The LLE can also be written in the frequency domain as a system of coupled mode equations (CMEs),
 \begin{equation}
 \begin{split}
 \frac{\partial a_\mu}{\partial t} ={}&  -\left ( \frac{\kappa}{2} +i\delta_\mu  \right) a_\mu + ig_0 \displaystyle\sum_{\mu',\mu''} {a_{\mu'} a_{\mu''} a^{*}_{\mu' + \mu''- \mu}} + \delta_{\mu,0}\sqrt{\kappa_\mathrm{ex}} s_\mathrm{in}  
    %t_R\frac{\partial A_\mu}{\partial \tau} = \left [ -\sigma -i\delta_0 - L\frac{i\beta_2}{2}\omega_\mu^2 \right] A + i\gamma L \displaystyle\sum_{\substack{\mu',\mu'', \mu''' \\ \mu = \mu' + \mu'' - \mu'''}}{A_{\mu'} A_{\mu''} A_{\mu'''}^*} + \sqrt{\theta} A_{in}
    \label{LLE_freq}
    \end{split}
\end{equation}
based on the modal expansion $A'(\phi, t) = \sum_{\mu}a_\mu (t)\,e^{-i\mu \phi}$, where $a_\mu$ describes the amplitude of the mode with resonance frequency $\omega_\mu$ in the rotating frame of $\omega_p+D_1\mu$ where $D_1=2\pi/T_\mathrm{R}$, and $\delta_\mu = \omega_0 + \tfrac{D_2}{2}\mu^2 - \omega_p$ is the mode detuning. %and $\delta_\mu = \omega_\mu - \omega_p - \mu D_1 = \delta_0 + D_{int} $ is the mode detuning defined in terms of the integrated dispersion $D_{int}$.

Our second simulation is based on the LLE/CME formulation, 
% framework combines the time formulation of the LLE and the frequency-domain set of CME. Specifically, the linear dynamics is computed using the modal formulation of Eq. (\ref{LLE_freq}), while the nonlinear Kerr interaction is evaluated in the time domain using the nonlinear term of Eq. (\ref{LLE}). The modal convolution is therefore replaced by an equivalent time-domain product, reducing the computational cost. The output field is then obtained as
and the output field is obtained via
\begin{equation}
    a_{out,\mu} = \delta_{\mu,0}s_{in} - \sqrt{\kappa_\mathrm{ex}}a_\mu.
\end{equation}

Both Ikeda-map and LLE/CME-based models can be extended to include additional nonlinear processes and noise sources in Kerr microresonators \cite{Hansson}, such as Raman self-frequency shift \cite{Raman}, thermorefractive noise \cite{Thermorefractive}, and pump phase and intensity noise \cite{Lei2022}. Here, we consider vacuum quantum noise
% In particular, we assume to have shot-noise-limited pumps $s_\text{in} = \bar{A}^{(')}_\text{in} + \delta A^{(')}_{\text{in}}$, consisting of a constant term $\bar{A}^{(')}_\text{in}$ proportional to the pump power and a fluctuation term $\delta A^{(')}_{\text{in}}$. 
which is included as an additional field coupled into the resonator and injected into each cavity mode, enabling the simulation of spontaneous light generation through FWM below the oscillation threshold. In the following, we describe the quantum noise implementation in the two semi-classical models, following the method described in Ref. \cite{Paschotta2004}.

In the Ikeda map, the noise field $\delta A_{\text{in}}$ coupled into the resonator with the classical pump is defined to have an autocorrelation function equal to
\begin{equation}
    \langle \delta A^*_{\text{in}}(t) \delta A_{\text{in}}(t + t') \rangle = \frac{\hbar \omega_p}{2}\delta(t')
    \label{auto}
\end{equation}
where $\hbar \omega_p$ is the photon energy. The noise term is actually defined in the frequency domain independently for each cavity mode simulated in the system as
\begin{equation}
    \delta \widetilde{A} _{\text{in},\mu} (\omega_\mu) = \left ( \eta_\mu + i\eta'_\mu \right)\frac{1}{\sqrt{2}} \sqrt{\frac{\hbar\omega_\mu}{2FSR}} 
    \label{quantum_noise}
\end{equation}
%normalized such that $|\widetilde{A} _{\text{in},\mu}|^2$ is the power in the frequency bin, 
where the terms $\eta_\mu$ and $\eta'_\mu$ are independent normally distributed real variables with zero mean and unit standard deviation. For each cavity roundtrip, Eq. (\ref{quantum_noise}) defines the noise contribution of each mode field by imposing the vacuum noise variances to be equal to the shot noise level, corresponding to an average of half a photon per frequency bin. The corresponding time-domain noise $\delta A_{\text{in}}(t)$ is introduced into Eqs. (\ref{Aout})-(\ref{An1}) with the same coupling coefficient as the input pump. In addition to the shot-noise-limited pump, we include an extra noise term related to the optical cavity loss, $\delta A_{\text{loss}}$, defined in the same way as Eq. (\ref{quantum_noise}). %The practical meaning of this term is to compensate for the vacuum photons lost in each roundtrip due to the intrinsic cavity loss, naturally described by Eq. (\ref{nlse}). It is therefore necessary to ensure that the intracavity noise level does not go below the shot noise level, as imposed by the Heisenberg uncertainty principle. 
This field is included in Eq. (\ref{An1}) with a coupling coefficient equal to the per-roundtrip propagation loss $\alpha L$. The complete Ikeda map input-output equations, including the noise terms, become
\begin{equation}
    A_\text{out} = A_\text{in}\sqrt{1-\theta} - A_n\sqrt{\theta} + \delta A_{\text{in}} \sqrt{1-\theta}
    \label{Aout2}
\end{equation}
\begin{equation}
    A_{n+1} = A_\text{in}\sqrt{\theta} + A_n\sqrt{1-\theta}  + \delta A_{\text{in}} \sqrt{\theta} +  \delta A_{\text{loss}} \sqrt{\alpha L}.
    \label{An12}
\end{equation}

The noise implementation in the CME-based simulation follows the same principle: vacuum fluctuations enter each mode through the bus waveguide and through the intrinsic loss channel, at the combined rate $\kappa$. \eqref{LLE_freq} becomes a stochastic differential equation, which we propagate numerically according to
\begin{equation}  
    \mathrm{d}a_\mu = \mathrm{RHS}_{\mathrm{\eqref{LLE_freq}}}\,\mathrm{d}t + \sqrt{\frac{\kappa}{2}}\,\mathrm{d}W_\mu ,
    \label{eq:stochastic_modal_lle}
\end{equation}
where $\mathrm{RHS}_{\mathrm{\eqref{LLE_freq}}}$ is the right-hand side of~\eqref{LLE_freq}, $\mathrm{d}t$ is the propagation time step, and $\mathrm{d}W_\mu$ are independent complex Wiener increments,
\begin{equation}
    \langle \mathrm{d}W_\mu \rangle = 0, \qquad
    \langle \mathrm{d}W^*_\mu\, \mathrm{d}W_\nu \rangle = \delta_{\mu\nu}\,\mathrm{d}t .
\end{equation}
In the prefactor, $\kappa$ is the total rate at which vacuum fluctuations enter the resonator, and the factor $1/2$ is their variance of half a photon per mode, consistent with~\eqref{auto}. Practically,~\eqref{eq:stochastic_modal_lle} can be integrated with the Euler--Maruyama scheme \cite{alma} (or similar method), where for each propagation step of duration $\mathrm{d}t$ we compute
\begin{equation}
    \mathrm{d}W_\mu = \sqrt{\mathrm{d}t}\;\frac{\varepsilon_\mu + i\varepsilon'_\mu}{\sqrt{2}},
\end{equation}
with $\varepsilon_\mu$, $\varepsilon'_\mu$ independent real Gaussian random variables of zero mean and unit variance (drawn new for every step).

\section{Simulation results}
Here, we present the simulation results obtained using the two semi-classical methods described above to investigate the noise properties of the processes in Fig. \ref{fig1}. In both simulation approaches, we use the same ring parameters, corresponding to realistic values for state-of-the-art SiN strongly overcoupled microresonators, with slight adjustments for each case study.  Finally, we benchmark the results against the analytical expressions derived in the corresponding quantum models, described in more detail in Supplement 1. 

\subsection{Two-mode quadrature squeezing}
\begin{figure*}[!]
\centering\includegraphics[]{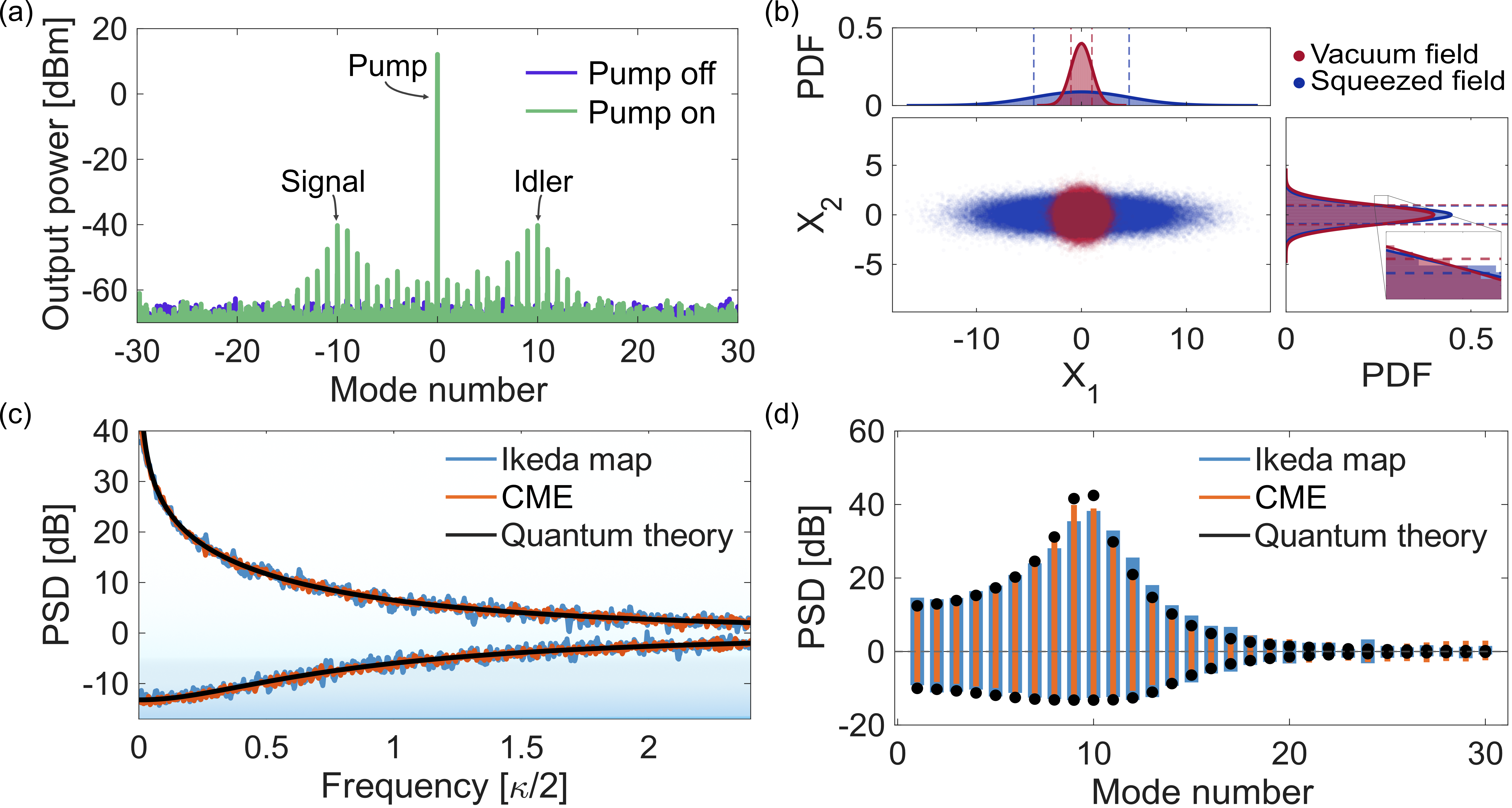}
\caption{Two-mode quadrature squeezing. (a) Output optical spectra obtained with pump on (green) and off (purple). The pump wavelength is set to $\lambda_p = \SI{1550}{\nano\meter}$, with $\delta_0 = \SI{0.0067}{\radian}$, $\beta_2 = -2.224\cdot10^{-25}\SI{}{\second^2/\meter}$, and $\text{FSR} = \SI{100}{\giga\hertz}$. (b) Phase space representation of the joint quadratures $X_1$ and $X_2$ of the primary modes $\mu = \pm10$ compared to the vacuum field, with the corresponding probability density functions (PDFs) and the standard deviations represented by the dashed lines. Panels (a) and (b) are simulated using the Ikeda map; similar results are obtained with the CME. (c) Squeezing and anti-squeezing noise spectra of the primary modes, normalized to the vacuum noise. PSD: power spectral density. (d) Mode number-dependent squeezing and anti-squeezing levels obtained for optimal $\theta_s$, evaluated at $\SI{1}{\mega\hertz}$. Panels (c) and (d) show results obtained with both the Ikeda map and the CME compared to the corresponding quantum theory.} 
\label{Fig2}
\end{figure*}
In this case, we pump the microresonator just below threshold, with $P_\text{in} = 99\% P_\text{th}$. To operate in the overcoupling regime, we set the ring-bus coupling rate $\kappa_\text{ex}/2\pi = \SI{200}{\mega\hertz}$ to exceed the intracavity loss rate $\kappa_0/2 \pi = \SI{10}{\mega\hertz}$. Fig. \ref{Fig2}(a) shows a portion of the corresponding output spectrum obtained from the Ikeda map-based simulation. Both approaches provide direct and simultaneous access to the time evolution of all signal and idler mode pairs. In the Ikeda map-based approach, the time evolution of the modes is obtained by storing the complex values of the individual frequency components obtained by Fourier transforming $A_\text{out}$ after each roundtrip, while in the CME-based approach, it is directly provided by Eq. (\ref{LLE_freq}).
%In the LLE-based approach, instead, Eq. (\ref{LLE_freq}) already provides the time evolution of the individual frequency modes. 

We first consider the two modes corresponding to the peaks of the parametric gain spectrum $\mu = \pm 10$. 
To investigate the squeezing, we combine the signal $A_s$ and idler $A_i$ output fields to calculate the joint quadrature \cite{Caves85}
\begin{equation}
    X(\theta_s) =\frac{1}{2\sqrt{2}} \left[ (A_s + A_i) e^{i\theta_s} + (A_s^* + A_i^*) e^{-i\theta_s}\right]
    \label{xtheta}
\end{equation}
where $A_{s(i)}^*$ is the complex conjugate field, and $\theta_s$ is the squeezing angle, corresponding experimentally to the phase of the local oscillator used to measure the squeezing. We first analyze the noise properties in the phase space, as shown in Fig. \ref{Fig2}(b). The combined signal and idler field distribution is compared to the vacuum field, obtained by repeating the same simulation with $P_\text{in} = \SI{0}{\watt}$, effectively introducing only vacuum noise into the system. As throughout this work, the field distributions are normalized such that the vacuum field has unitary standard deviations. The joint quadratures $X_1$ and $X_2$ are obtained from Eq. (\ref{xtheta}) with a relative $\pi/2$ phase shift. The simulation results show that the noise of $X_1$ is significantly amplified above the vacuum noise while the noise of $X_2$ is reduced, demonstrating two-mode quadrature squeezing. Here, the squeezing angle $\theta_s$ determines the orientation of the noise ellipse and is optimized such that $X_2$ is squeezed while $X_1$ is anti-squeezed.

We quantify the squeezing through the corresponding noise spectrum in the Fourier domain. Figure \ref{Fig2}(c) shows the squeezing and anti-squeezing spectra obtained from both semi-classical simulations and the theoretical quantum predictions based on Refs. \cite{Chembo2016, steady}.
% \begin{equation}
%     S_\text{sq}(\Omega) = 1 - \rho \frac{4 \kappa_a^2}{\Omega^2 + 4\kappa_a^2}
%     \label{sqTWQSS}
% \end{equation}
% \begin{equation}
%     S_\text{asq}(\Omega) = 1 + \rho \frac{4 \kappa_a^2}{\Omega^2} \left [ 1 + \frac{4 \kappa_p^2}{\Omega^2 + 4\kappa_a^2} \right ]^2
%     \label{asqTWQSS}
% \end{equation}
% with linear coefficients
% \begin{equation}
%     \kappa_a = -g_0N_p \sin{\left(2\phi_p - 2\phi_{s(i)}\right)}
%     \label{ka}
% \end{equation}
% \begin{equation}
%     \kappa_p = g_0 \left[N_{s(i)} +  N_p\cos{\left(2\phi_p - 2\phi_{s(i)}\right)}\right].
%     \label{kp}
% \end{equation}
% Here, $\Omega$ is the detection angular frequency, $\rho = \frac{\kappa_e}{\kappa}$ is the fraction of the coupling rate $\kappa_e$ over the total decay rate $\kappa = \kappa_i + \kappa_e$, $g_0 = \hbar \omega_p v_g^2 \gamma/L$ is the nonlinear coupling coefficient, where $v_g$ is the group velocity, and $\phi_{p(s,i)}$ and $N_{p(s,i)}=dt |A_{p(s,i)}|^2/(\hbar \omega_{p(s,i)})$ are, respectively, the phase and the photon number of the pump (signal and idler) mode, with $dt$ the time interval between subsequent samples of $A_{p(s,i)}$. 
Both semi-classical approaches give consistent results and agree with the quantum model. 

%Here, $\Omega$ is the detection angular frequency, $\rho = \frac{\kappa_e}{\kappa}$ is the fraction of the coupling rate $\kappa_e$ over the total cavity decay rate $\kappa = \kappa_i + \kappa_e$, $|\mathcal{A}_{p(s,i)}|^2$ and $\phi_{p(s,i)}$ are respectively the photon number and the phase of the pump (signal, and idler) mode, and $g_0 = \hbar \omega_p v_g^2 \gamma/L$ is the nonlinear coupling coefficient, where $v_g$ is the group velocity. Our results are in strong agreement with the quantum models, validating our approaches. 

%$N_s = dt |A_s|^2/(\hbar \omega_s)$ and $N_i = dt |A_i|^2/(\hbar \omega_i)$ respectively, where $dt$ is the time interval between subsequent samples of $A_{s(i)}$. 

We repeat the same analysis for all modes in the quantum frequency comb and report the squeezing and anti-squeezing levels obtained at the optimal $\theta_s$ in Fig. \ref{Fig2}(d). As expected, the largest squeezing and anti-squeezing levels are achieved with the strongest modes $\mu = \pm10$. These numerical results again show good agreement with the prediction of the quantum theory, for which the optimal squeezing and anti-squeezing levels are obtained from the analytical expressions derived in Ref. \cite{steady}. 
% \begin{equation}
% \begin{split}
%     S^\text{best}_\text{sq(asq)} (\mu) ={}& 1 + \frac{8 \eta \kappa_e \kappa^2}{4 \Delta_{\mu,\text{eff}}^2 + \kappa^2 \left( 1-\frac{P_\text{in}}{P_\text{th}} \right)}  + \\& \mp 4 \left |  \sqrt{\frac{\eta \kappa_e \kappa P_\text{in} P_\text{th} \left[  \kappa^2 P_\text{in}^2 - P_\text{th}^2 (2\Delta_{\mu,\text{eff}} + i\kappa )^2 \right]}{\left[  4\Delta_{\mu,\text{eff}}^2 P_\text{th}^2 + \kappa^2 ( P_\text{th}^2 - P_\text{in}^2  ) \right ]^2 }}  \right |^2
%  \end{split}
% \end{equation}
% where $\eta$ is the post-coupling efficiency, here assumed to be $\eta = 1$, and $\Delta_{\mu,\text{eff}} = \frac{1}{2} D_2 \mu^2 - \frac{\kappa}{2}P_\text{in}/{P_\text{th}}$ is the effective mode detuning, which includes the effect of the second-order dispersion defined as $D_2 = -\text{FSR}^3L\beta_2(2\pi)^2$.

\subsection{Single-mode quadrature squeezing}
\begin{figure}[!t]
\centering\includegraphics[]{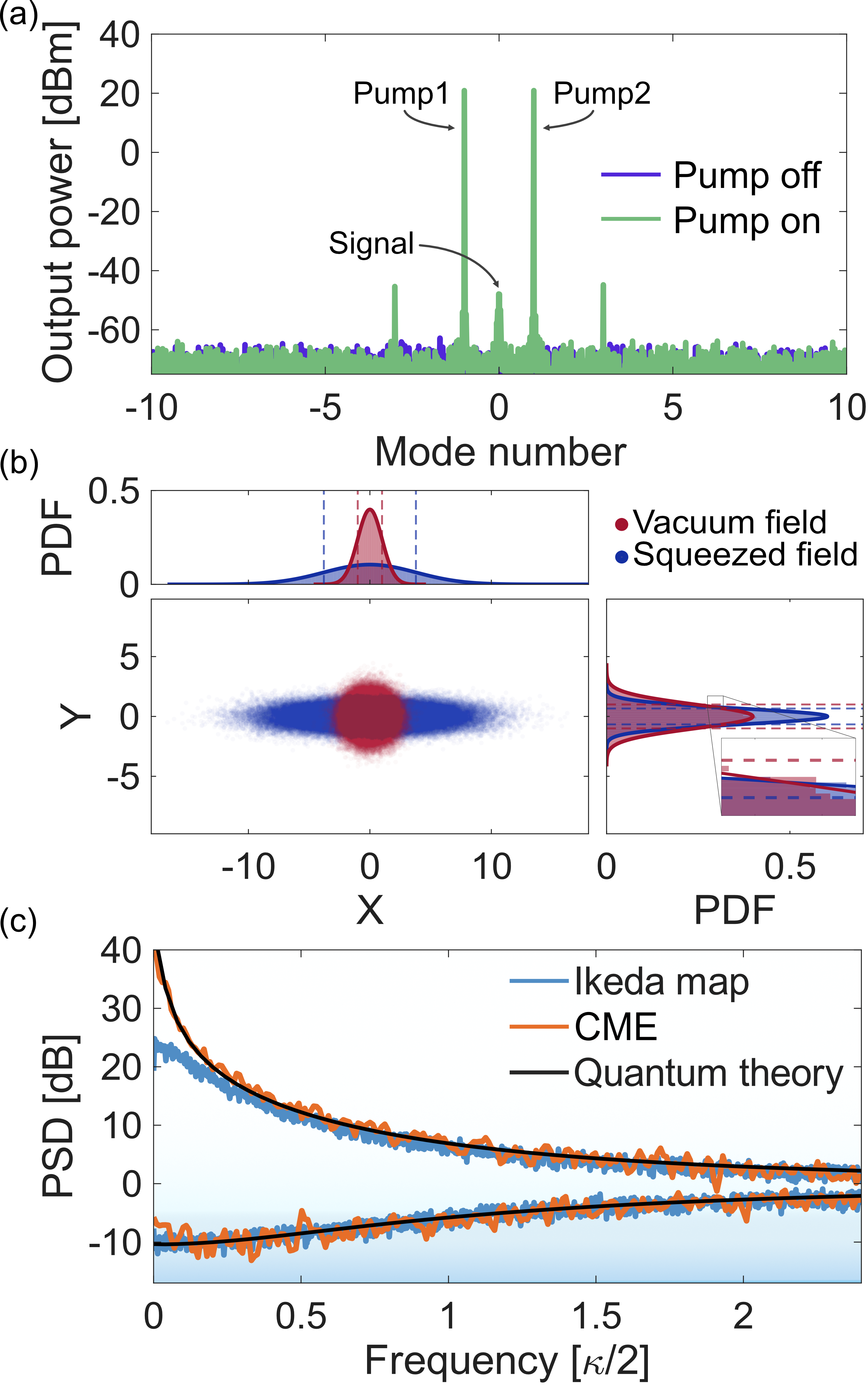}
\caption{Single-mode quadrature squeezing. (a) Output optical spectra obtained with pumps on (green) and off (purple). The pump wavelengths, separated by 2 FSRs, are $\lambda_\text{p1} = \SI{1550.8}{\nano\meter}$ and $\lambda_\text{p2} = \SI{1549.2}{\nano\meter}$, with $\delta_0 = \SI{0.044}{\radian}$, $\beta_2 = \SI{0}{\second^2/\meter}$, $\text{FSR} = \SI{100}{\giga\hertz}$, $\kappa_0/2\pi = \SI{80}{\mega\hertz}$, and $\kappa_\text{ex}/2\pi = \SI{800}{\mega\hertz}$. (b) Phase space representation of the field quadratures $X$ and $Y$ of the central signal mode compared to the vacuum field, together with the corresponding PDF estimates and the standard deviations shown by the dashed lines. Panels (a) and (b) are simulated using the Ikeda map. Similar results are obtained with the CME-based approach. (c) Squeezing and anti-squeezing noise spectra of the signal normalized to the vacuum noise, simulated using both the Ikeda map and the CME, and compared to the corresponding quantum theory.}
\label{fig3}
\end{figure}
In experiments, high levels of single-mode squeezing are achieved by suppressing the modes involved in parasitic processes~\cite{Zhang2021, Viola2024, Ulanov2025}. To emulate this situation, we restrict the CME-based approach to simulate only three modes: the two pumps and the central degenerate signal. Similarly, in the Ikeda map-based approach, we increase the effective detuning of all other modes so that only these three modes are resonant in the cavity. This simplified three-mode system is characterized by a parametric threshold above which the signal begins to oscillate \cite{He}. We set both pump powers slightly below this threshold. An example of the resulting output spectrum from the Ikeda map simulations is shown in Fig. \ref{fig3}(a). 

To analyze the squeezing, we calculate the quadrature of the signal mode in the same way as in Eq. (\ref{xtheta}), considering the signal and idler modes to be degenerate, i.e.
\begin{equation}
    X_s(\theta_s) = \frac{1}{\sqrt{2}} \left[ A_s  e^{i\theta_s} + A_s^* e^{-i\theta_s}\right].
    \label{xtheta_deg}
\end{equation}
The resulting phase space distribution is shown in Fig. \ref{fig3}(b), together with the corresponding vacuum field. As in the previous case, the quadratures $X$ and $Y$ are obtained from Eq. (\ref{xtheta_deg}) with optimal $\theta_s$ and a relative $\pi/2$ phase shift. As expected, the distribution exhibits the typical noise ellipse of a squeezed state. 
We further study the squeezing and anti-squeezing noise spectra, shown in Fig. \ref{fig3}(c), and compare them with the predictions from the quantum model of the three-mode system. We obtain the latter by following the derivation in Refs. \cite{Chembo2016, Guidry2023, gouzien:2020}.
% where the squeezing spectrum is derived from the numerical evaluation of the output signal quadrature vector 
% \begin{equation}
%     \textbf{Q}_\text{out}(\omega) = \sqrt{\kappa_e} \left (i\omega \mathbb{I} + \textbf{M}_q \right )^{-1} \left(\sqrt{\kappa_i}\textbf{U} + \sqrt{\kappa_e}\textbf{V} \right) + \textbf{V}
% \end{equation}
% where $\textbf{U}$ and $\mathbf{V}$ are the quadrature vectors of the vacuum fields, and $\textbf{M}_q$ is a matrix describing the linear and nonlinear processes. 
Also in this case, our results are consistent with the theoretical predictions.

\subsection{Intensity difference squeezing}
\begin{figure}[!t]
\centering\includegraphics[]{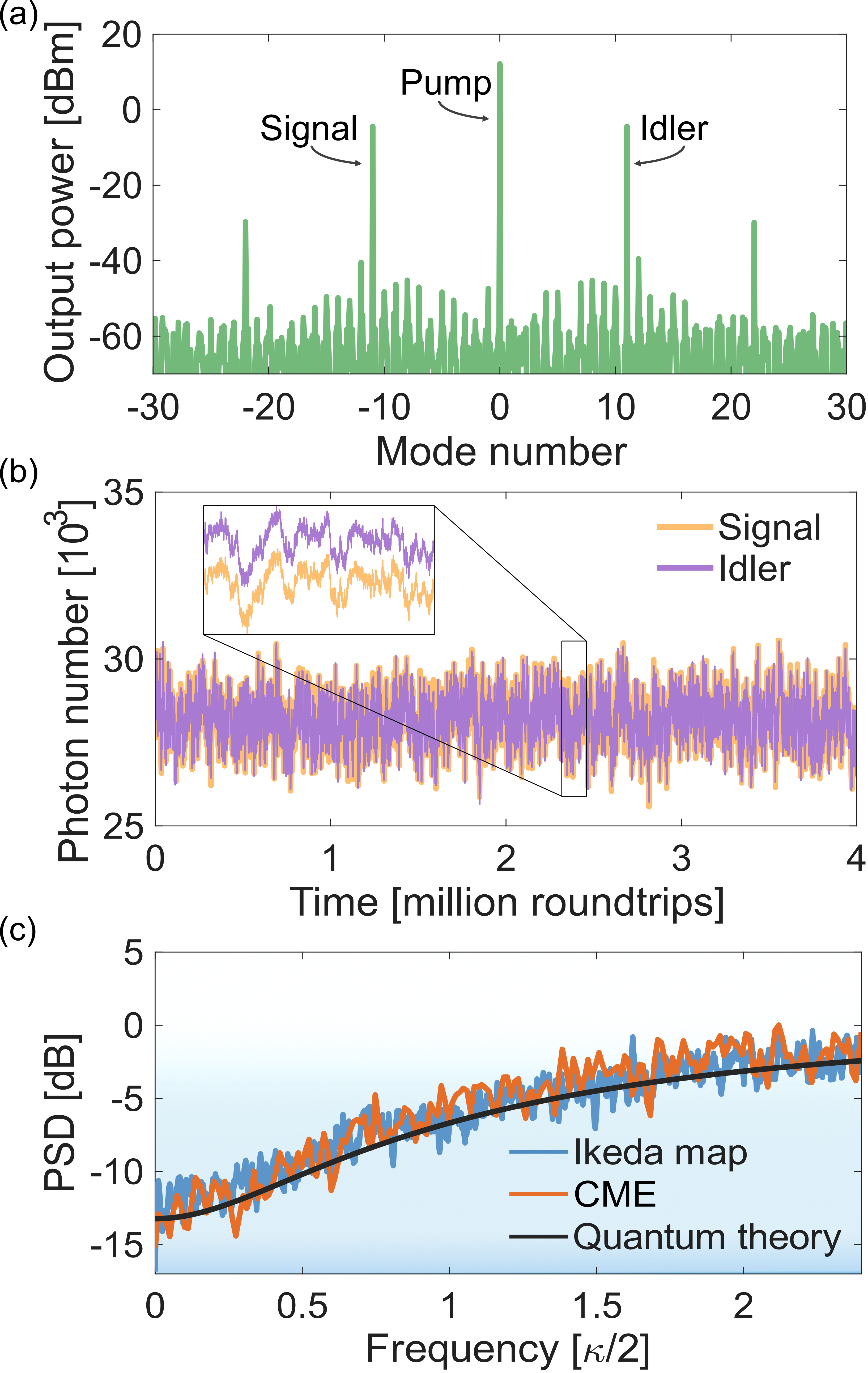}
\caption{Intensity difference squeezing. (a) Output optical spectrum. The ring parameters are the same as for Fig. \ref{Fig2}, except for the pump ratio set to $P_\text{in}/P_\text{th} = 103\%$. (b) Photon number fluctuations of the bright signal and idler modes, displaced in the inset to better show the correlations. Panels (a) and (b) are simulated using the Ikeda map, obtaining similar results also from the CME. (c) Squeezing noise spectrum of the bright modes normalized to the shot noise, simulated using both the Ikeda map and the CME, and compared to the corresponding quantum theory.}
\label{fig4}
\end{figure}

The last type of squeezing investigated is intensity difference squeezing. In this case, the microresonator has the same parameters used to generate two-mode quadrature squeezing, except that the input pump power is set above threshold $P_\text{in} = 103\%P_\text{th}$. As a consequence, the output optical spectrum, shown in Fig. \ref{fig4}(a), is characterized by bright signal and idler mode pairs. Here, we focus our analysis on the first bright pair closer to the pump, corresponding to modes $\mu = \pm10$. %Figure \ref{fig4}(b) shows the strong correlation between the photon numbers of these modes % defined over many cavity roundtrips. % as $N_s = dt |A_s|^2/(\hbar \omega_s)$ and $N_i = dt |A_i|^2/(\hbar \omega_i)$ respectively, where $dt$ is the time interval between subsequent samples of $A_{s(i)}$. 

As described previously, this type of squeezing arises from the correlations between the signal and idler photon numbers, $N_s$ and $N_i$. Figure \ref{fig4}(b) shows these correlations obtained after many cavity roundtrips from the Ikeda map-based simulations.
The squeezing spectrum is then analyzed by evaluating the power spectrum of the photon number difference $N_\text{diff} = N_s - N_i$ normalized to the shot noise level defined by the photon number sum of two coherent fields with the same signal and idler powers \cite{Chekhova2015, Chembo2016}. The squeezing spectra resulting from our simulation approaches are shown in Fig. \ref{fig4}(c) and compared with the quantum predictions based on Ref. \cite{Chembo2016}. 
% \begin{equation}
%     S'_\text{sq}(\Omega) = 1 - \rho \frac{4 \kappa^2}{\Omega^2 + 4\kappa^2}.
% \end{equation}
As in the previous cases, the results of our simulations are in agreement with the quantum theory.  %In this case, the anti-squeezed observable is the photon number sum of the bright signal and idler modes $N_\text{sum} = N_s + N_i$. 

\section{Summary and discussion}
In summary, we demonstrate that semi-classical models based on the Ikeda map and on the CME, typically used to investigate the classical nonlinear behavior of Kerr microresonators, can simulate squeezing generation when correctly supplemented with vacuum noise. We validate our approaches by investigating the noise properties of microresonator modes under three pump configurations, operating it above or below the optical threshold with either a single or dual input pump, to generate two-mode and single-mode quadrature squeezing and intensity difference squeezing. We use the typical parameters of a SiN overcoupled microresonator, however, they can be easily adapted to other materials or more complex resonator designs, such as photonic molecules \cite{Dutt:16, Zhang2021} or photonic crystal resonators \cite{Lu2022, Ulanov2025}. The squeezing and anti-squeezing properties of the cavity modes are investigated in both the phase space and noise spectra. 

In all cases, our simulation results agree with the corresponding quantum theoretical description, demonstrating that semi-classical models can predict the noise squeezing properties in nonlinear microresonators by effectively simulating the phase-sensitive amplification \cite{PSA} of the vacuum fluctuations. Our results suggest that the simulations can be leveraged to describe cases where analytic models are challenging to apply. For instance this includes broadband spectra, multi-resonator systems, as well as, studying the impact of additional effect such as Raman scattering, thermorefractive, or technical noise. 

% complex Kerr combs, such as multimode squeezing in Turing-roll patterns or soliton states, chaotic comb states, parasitic processes, or additional effects such as Raman scattering or thermorefractive noise, as well as in coupled resonator geometries such as photonic molecules \cite{Dutt:16} or photonic crystal resonators \cite{Lu2022}.
% Moreover, these semi-classical formalisms can be extended to include additional nonlinear effects or unwanted noise sources and interrogate their influence on the achievable squeezing levels. They can be used to investigate squeezing generation in more complex Kerr combs, such as multimode squeezing in Turing-roll patterns or soliton states, contributing to the development of efficient and application-optimized integrated squeezers. 
Semi-classical models therefore provide powerful tools for further investigating the squeezing properties of microresonators, complementing the quantum models in designing efficient squeezed sources. 
%This makes them powerful tools for designing efficient squeezed sources, complementing the existing quantum models.

\begin{backmatter}
\bmsection{Funding}VR excellence center (2022-0657); Swedish Research Council (VR starting grant 2023-0519); National Natural Science Foundation of China (NSFC) (62475042); The 111 project (B25030);  European Innovation Council (grant agreement No 101137000); Helmholtz VH-NG-1404; V.B.A. and V.T.C. acknowledge the Wallenberg Center for Quantum Technology (WACQT) in Sweden; V.B.A. acknowledges the support from the Göran Gustafsson Prize for early career researchers.

% For preprint submissions, please include funder names and grant numbers in the manuscript.

%\bmsection{Acknowledgment}V.B.A. acknowledges the support from the Göran Gustafsson Prize for early career researchers and the Wallenberg Center for Quantum Technology (WACQT) in Sweden.

\bmsection{Disclosures}The authors declare no conflicts of interest.

\bmsection{Data availability}Data underlying the results presented in this paper are available in Ref. (not yet created). 

\bmsection{Supplemental document}See Supplement 1 for supporting content.

\end{backmatter}

%%%%%%%%%% If using BibTeX:
\bibliography{sample}

\end{document}